\documentclass{article}
\usepackage[english]{babel} 

\usepackage{graphicx}

\usepackage{color}

\begin{document}

\newcommand{\rum}{\rule{0.5pt}{0pt}}
\newcommand{\rub}{\rule{1pt}{0pt}}
\newcommand{\rim}{\rule{0.3pt}{0pt}}
\newcommand{\numtimes}{\mbox{\raisebox{1.5pt}{${\scriptscriptstyle \rum\times}$}}}
\newcommand{\numtimess}{\mbox{\raisebox{1.0pt}{${\scriptscriptstyle \rum\times}$}}}
\newcommand{\Boldsq}{\vbox{\hrule height 0.7pt
\hbox{\vrule width 0.7pt \phantom{\footnotesize T}%
\vrule width 0.7pt}\hrule height 0.7pt}}
\newcommand{\two}{$\raise.5ex\hbox{$\scriptstyle 1$}\kern-.1em/
\kern-.15em\lower.25ex\hbox{$\scriptstyle 2$}$}

\renewcommand{\refname}{References}
\renewcommand{\tablename}{\small Table}
\renewcommand{\figurename}{\small Fig.}
\renewcommand{\contentsname}{Contents}

\begin{center}
{\Large\bf 
Dynamical 3-Space:  Cosmic Filaments, Sheets and Voids\rule{0pt}{13pt}}\par

\bigskip
Reginald T. Cahill \\ 
{\small\it  School of Chemical and Physical  Sciences, Flinders University,
Adelaide 5001, Australia\rule{0pt}{15pt}}\\
\raisebox{+1pt}{\footnotesize E-mail: Reg.Cahill@flinders.edu.au}\par

\bigskip

{\small\parbox{11cm}{%
Observations of weak gravitational lensing combined with  statistical tomographic techniques have revealed that galaxies have formed along filaments, essentially one-dimensional lines or strings, which form sheets and voids.  These have, in the main,  been interpreted as ``dark matter" effects.   To the contrary here we report the discovery that the dynamical 3-space theory possesses such filamentary solutions.  These solutions are purely space self-interaction effects, and are  attractive to matter, and as well generate electromagnetic lensing.  This  theory of space has explained bore hole anomalies, supermassive black hole masses in spherical galaxies and globular clusters,  flat rotation curves of spiral galaxies, and other gravitational anomalies. The theory has two constants, $G$ and $\alpha$, where the bore hole experiments show that $\alpha \approx 1/137$ is the fine structure constant.
\rule[0pt]{0pt}{0pt}}}\medskip
\end{center}

\setcounter{section}{0}
\setcounter{equation}{0}
\setcounter{figure}{0}
\setcounter{table}{0}

\markboth{Cahill R.T.  Dynamical 3-Space:  Cosmic Filaments, Sheets and Voids}{\thepage}
\markright{ Cahill R.T.  Dynamical 3-Space:  Cosmic Filaments, Sheets and Voids}

\section{Introduction}

\begin{figure}[th]
\hspace{30mm}\vspace{0mm}\hspace{10mm}\includegraphics[scale=2.0]{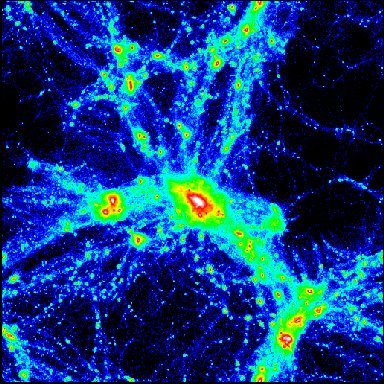}

\hspace{42mm}\rotatebox{90}{\includegraphics[scale=0.4]{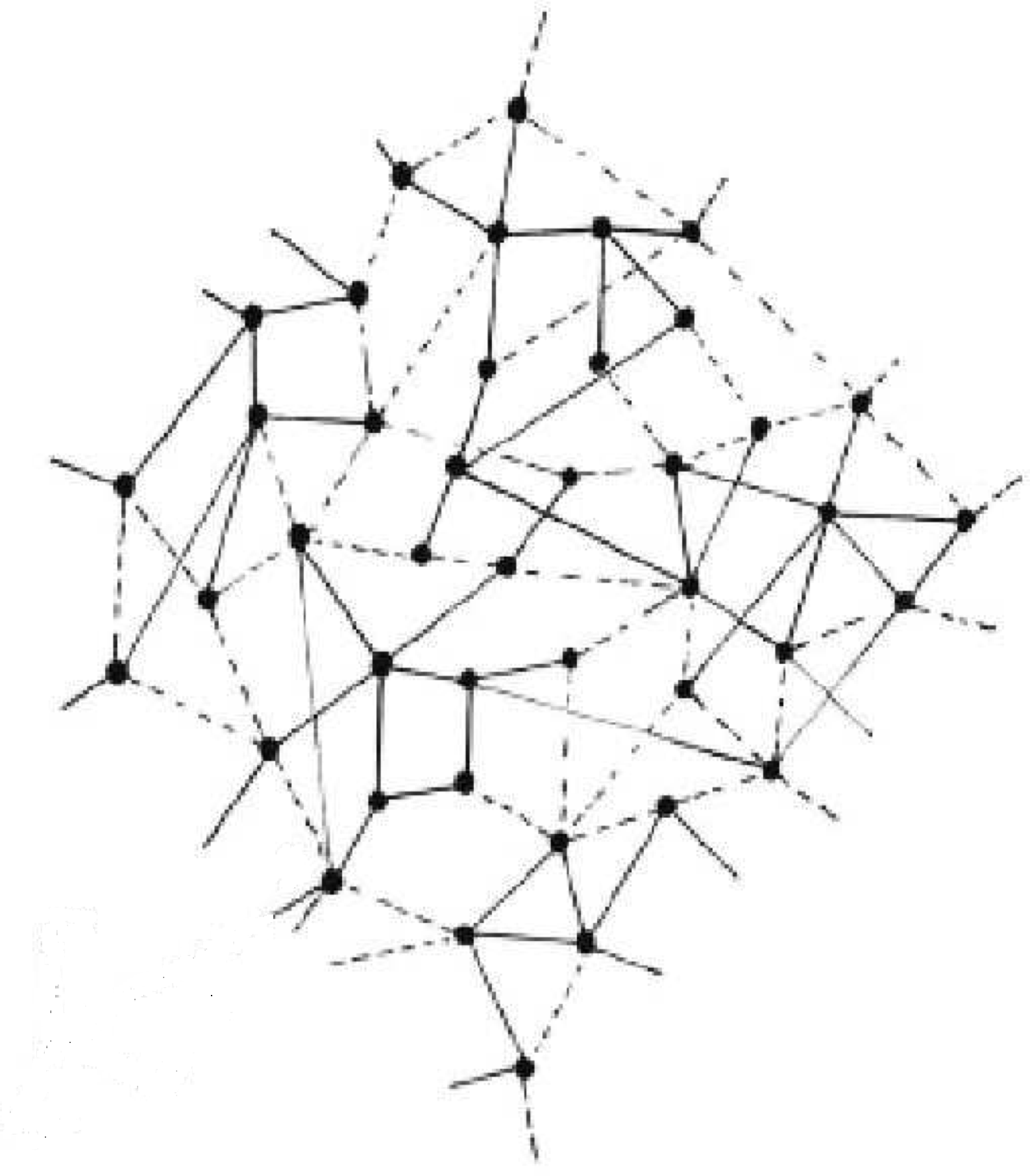} }
\vspace{-4mm}\caption{\small{Top: Cosmic filaments as revealed by gravitational lensing statistical tomography. From J.A.  Tyson and G. Bernstein, Bell Laboratories, Physical Sciences Research, http://www.bell-labs.com/org/physicalsciences/projects/darkmatter/darkmatter.html. Bottom:  Cosmic network of primordial filaments and primordial black holes, as solution from (\ref{eqn:3spacedynamics}).}}
\label{fig:Filaments}
\end{figure}

Observations of weak gravitational lensing and statistical tomographic techniques have revealed that galaxies have formed along filaments, essentially one-dimensional lines or strings  \cite{Clusters}, see Fig.\ref{fig:Filaments}. These have, in the main,  been interpreted as ``dark matter" effects.  Here we report the discovery that the dynamical 3-space theory possesses such filamentary solutions, and so does away with  the ``dark matter" interpretation.  The dynamical 3-space theory is a uniquely determined generalisation of Newtonian gravity, when that is expressed in terms of a velocity field, instead of the original gravitational acceleration field \cite{Book,Review}.  This velocity field has been repeatedly detected via numerous light speed anisotropy experiments, beginning with the 1887 Michelson-Morley gas-mode interferometer experiment \cite{MMCK, MMC}.  This is a theory of space, and has explained bore hole anomalies, supermassive black hole masses in spherical galaxies and globular clusters,  flat rotation curves of spiral galaxies, and other gravitational anomalies. The theory has two constants, $G$ and $\alpha$, where the bore hole experiments show that $\alpha \approx 1/137$ is the fine structure constant. The filamentary  solutions are purely a consequence of the space self-interaction dynamics, and are  attractive to matter, and as well generate electromagnetic lensing.  The same self-interaction dynamics has been shown to generate inflow singularities, {\it viz} black holes  \cite{CahillBH2}, with both the filaments and black holes generating   long-range non-Newtonian  gravitational forces. The dynamical 3-space also has Hubble expanding universe solutions that give a parameter-free account of the supernova redshift-magnitude data, without the need for  ``dark matter" or  "dark energy"  \cite{Paradigm}.   The black hole and filament solutions are primordial remnants of the big bang in the epoch when  space was self-organising, and then provided a framework for the precocious clumping of matter, as these inflow singularities are long-range gravitational attractors.  That $\alpha$ determines the strength of these phenomena implies that we are seeing evidence of a unification of space, gravity and quantum theory, as conjectured in Process Physics \cite{Book}.

\section{Dynamical 3-Space}
The dynamics of space is easily  determined by  returning to Galileo's discoveries of the free-fall acceleration of test masses, and  using a velocity field to construct a minimal and unique formulation  that  determines the acceleration of space itself
\cite{Book,EmergentGravity}. In the case of zero vorticity  we find 
 \begin{equation}
\nabla\! \cdot\!\left(\frac{\partial {\bf v} }{\partial t}+ ({\bf v}\!\cdot\!{\bf \nabla}){\bf v}\right)+
\frac{\alpha}{8}\left((tr D)^2 -tr(D^2)\right)+... =-4\pi G\rho \nonumber \label{eqn:3space}\end{equation}
\begin{equation}
 \nabla\times {\bf v}={\bf 0},  \mbox{\  \  \   }
 D_{ij}=\frac{1}{2}\left(\frac{\partial v_i}{\partial x_j}+
\frac{\partial v_j}{\partial x_i}\right),
\label{eqn:3spacedynamics}\end{equation} 
$G$ is Newton's constant, which has been revealed as  determining the dissipative flow of space into matter, and $\alpha$ is a dimensionless constant, that experiment reveals to be the fine structure constant.
The space acceleration is determined by the Euler constituent acceleration
\begin{equation}
{\bf a}=\displaystyle{\frac{\partial {\bf v}}{\partial t}}+({\bf v}\cdot{\bf \nabla}){\bf v}
\label{eqn:spaceaccel}\end{equation}
 The matter acceleration is found by determining the trajectory of a quantum matter wavepacket  to be \cite{Schrod}
\begin{equation}
{\bf g}=\displaystyle{\frac{\partial {\bf v}}{\partial t}}+({\bf v}\cdot{\bf \nabla}){\bf
v}+({\bf \nabla}\times{\bf v})\times{\bf v}_R-\frac{{\bf
v}_R}{1-\displaystyle{\frac{{\bf v}_R^2}{c^2}}}
\frac{1}{2}\frac{d}{dt}\left(\frac{{\bf v}_R^2}{c^2}\right)+...
\label{eqn:acceleration}\end{equation}
where ${\bf v}({\bf r}, t)$ is the velocity of a structured element of space wrt to an observer's arbitrary Euclidean coordinate system, but which has no ontological meaning.  The relativistic term in (\ref{eqn:acceleration}) follows  from extremising  the elapsed proper time wrt a quantum matter  wave-packet trajectory ${\bf r}_o(t)$, see \cite{Book}. This ensures that quantum waves propagating along neighbouring paths are in phase. 
\begin{equation}
\tau=\int dt \sqrt{1-\frac{{\bf v}^2_R({\bf r}_0(t),t)}{c^2}}
\label{eqn:propertime}\end{equation}
where ${\bf v}_R({\bf r}_o(t),t) ={\bf v}_o(t) - {\bf v}({\bf r}_o(t),t),$ is the velocity of the wave packet, at position ${\bf r}_0(t)$,  wrt the local 3-space, and ${\bf g}=d{\bf r}_O/dt$.  This shows that (i) the matter `gravitational' geodesic is a quantum wave refraction effect, with the trajectory determined by a Fermat maximum proper-time  principle, and (ii) that quantum systems undergo a local time dilation effect caused by their absolute motion wrt space. The last term in (\ref{eqn:acceleration}) causes the precession of planetary orbits.

It is essential that we briefly review some of the many tests that have been applied to this dynamical 3-space.

\subsection{Direct Observation of 3-Space}

 \begin{figure}[t]
\vspace{-0mm}
\hspace{35mm}\includegraphics[scale=0.8]{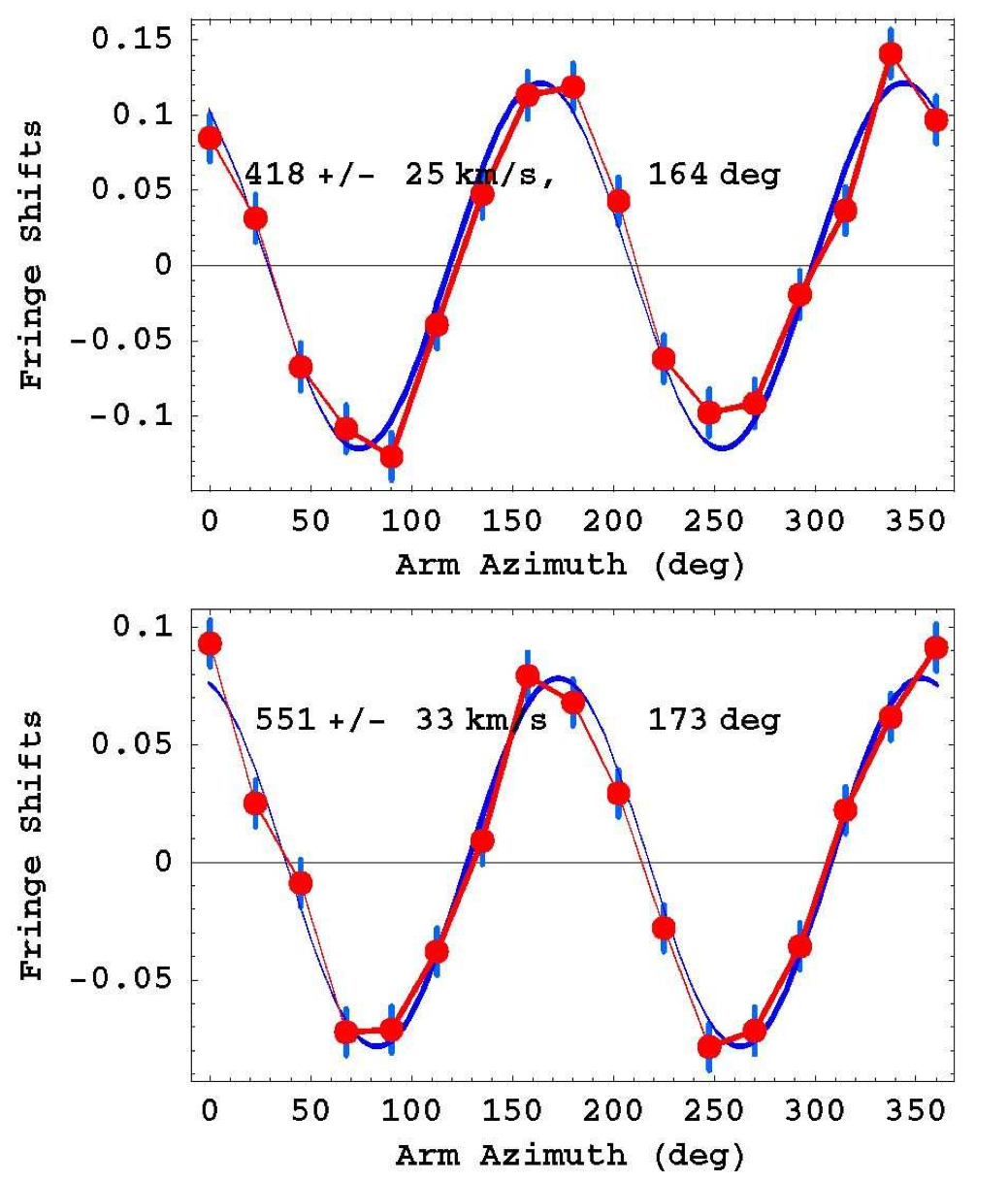}
\caption{\small {(a) A typical  Miller  averaged-data from September 16, 1925, $4^h 40^\prime$  Local Sidereal Time (LST) - an average of data from 20 turns of the gas-mode Michelson interferometer. Plot and data after fitting and then subtracting both the temperature drift and Hicks effects from both, leaving the expected sinusoidal form.  The error bars are determined as the rms error in this fitting procedure, and show how exceptionally small were the errors, and which agree with Miller's claim for the errors. (b) Best result  from the Michelson-Morley 1887 data - an average of 6 turns, at  $7^h$  LST on July  11, 1887.  Again the rms error is remarkably small.  In both cases the indicated speed is  $v_P$ - the 3-space speed projected onto the plane of the interferometer. The angle is  the azimuth of the  3-space speed projection at the particular LST.  The speed fluctuations from day to day significantly exceed these errors, and reveal the existence of 3-space flow turbulence - i.e gravitational waves.}}
\label{fig:MillerMMPlots}\end{figure}

Numerous direct observations of 3-space involve the detection of light speed anisotropy. These began with the 1887 Michelson-Morley gas-mode interferometer experiment, that gives a solar system galactic speed in excess of 300 km/s, \cite{MMCK,MMC}\footnote{Amazingly it continues to be claimed that this experiment was null.}. These experiments have revealed components of the flow, a dissipative inflow, caused by the sun and the earth, as well as the orbital motion of the earth.  The largest effect is the galactic  velocity of the solar system of 486 km/s in the direction RA = $4.3^\circ$,  Dec = $-75^\circ$,  determined from spacecraft earth-flyby Doppler shift data \cite{CahillNASA}, a direction first detected by Miller in his 1925/26 gas-mode Michelson interferometer experiment \cite{Miller}.

\subsection{Newtonian Gravity Limit}
In the limit of zero vorticity and neglecting relativistic effects  (\ref{eqn:3spacedynamics}) and (\ref{eqn:acceleration})
give\begin{equation}
\nabla\cdot{\bf g}=-4\pi G\rho-4\pi G\rho_{DM}, \mbox{\ \  } \nabla \times {\bf g}={\bf 0}
\label{eqn:NGplus}\end{equation}
where
\begin{equation}
\rho_{DM} =\frac{\alpha}{32\pi G}\left((tr D)^2 -tr(D^2)\right).
\label{eqn:darkmatter}\end{equation}
This is Newtonian gravity, but with the extra dynamical term which has been used to define an effective ``dark matter'' density.  This is not necessarily non-negative, so in some circumstances ant-gravity effects are possible, though not discussed herein.This $\rho_{DM}$  is not  a real matter density, of any form, but is the matter density needed within Newtonian gravity to explain dynamical effects caused by the $\alpha$-term in (\ref{eqn:3spacedynamics}). This term explains the flat rotation curves of spiral galaxies, large light bending and lensing effects from galaxies, and other effects.  However, it is purely a space self-interaction effect.

\subsection{Curved Spacetime Formalism}
Eqn.(\ref{eqn:propertime}) for the elapsed proper time maybe written
\begin{equation}
d\tau^2=dt^2\!-\!\frac{1}{c^2}(d{\bf r}(t)-{\bf v}({\bf r}(t),t)dt)^2=g_{\mu\nu}(x)dx^\mu dx^\nu,
\label{eqn:PGmetric}\end{equation}
which introduces a curved spacetime metric $g_{\mu\nu}$. However this spacetime has no ontological significance - it is merely a mathematical artifact, and as such hides the underlying dynamical 3-space. Its only role is to describe the geodesic of the matter quantum wave-packet in gerneral coordinates.  The metric is determined by solutions of (\ref{eqn:3spacedynamics}). This induced metric is not determined by the Einstein-Hilbert equations, which originated as a generalisation of Newtonian gravity, but without the knowledge that a dynamical 3-space had indeed been detected by Michelson and Morley in 1887 by detecting light speed anisotropy.

\subsection{Gravitational Waves}
Eqn.(\ref{eqn:3spacedynamics}) predicts time dependent  flows, and these have been repeatedly detected, beginning with the Michelson and Morley experiment in 1887. Apart from the sidereal earth-rotation induced time-dependence, the light-speed anisotropy data has always shown time-dependnet fluctuations/turbulence, and at a scale of some 10\% of the background galactic flow speed.  This  time dependent velocity field induces ``ripples''  in the spacetime metric in  (\ref{eqn:PGmetric}), which are known as ``gravitational waves".  They cannot be detected by a vacuum-mode Michelson interferometer.

\subsection{Matter Induced Minimal Black Holes}
\begin{figure}[t]
\hspace{30mm}\includegraphics[scale=0.25]{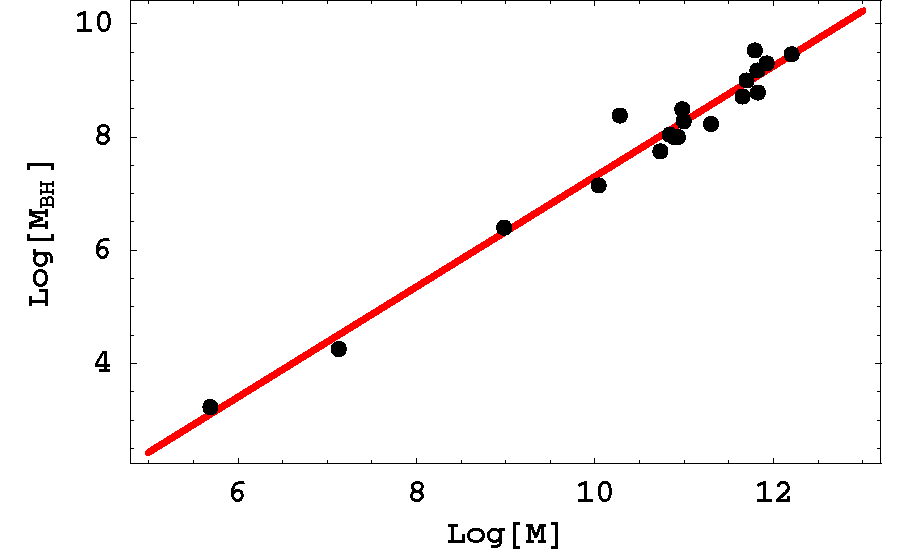}
\caption{\small{The data shows $\mbox{Log}_{10}[M_{BH}]$ for the minimal induced black hole masses $M_{BH}$  for
a variety of spherical matter systems, from Milky Way globular clusters to spherical galaxies, with masses $M$, plotted against  $\mbox{Log}_{10}[M]$, in solar masses $M_0$.  The straight line is the prediction from (\ref{eqn:BHmass}) with $\alpha=1/137$. See \cite{CahillBH2} for references to the data. }
\label{fig:BlackHoles}}\end{figure}

For  the special case of a spherically symmetric flow  we set  ${\bf v}({ \bf r}, t)=-{\bf {\hat r}} v(r,t)$.  Then (\ref{eqn:3spacedynamics}) becomes, with $v^\prime=\partial v/\partial r$,
\begin{equation}
\frac{\partial v^\prime}{\partial t}+v v''+\frac{2}{r} v v' + (v')^2 + \frac{\alpha}{2 r}\left(\frac{v^2}{2r}+v v' \right) =-4 \pi G \rho
\label{eqn:sphericalsym}\end{equation}
For a  matter density $\rho(r)$, with maximum radius $R$,  (\ref{eqn:sphericalsym}) has an exact inhomogeneous static solution \cite{Sun}
\begin{equation}
v(r)^2=\left\{
\begin{array}{l} 
     \displaystyle  \frac{2G}{(1-\frac{\alpha}{2})r} \int_0^r 4 \pi s^2 \rho(s) ds \smallskip \\ 
     \displaystyle  + \frac{2G}{(1-\frac{\alpha}{2})r^\frac{\alpha}{2}} \int_r^R 4 \pi s^{1+\frac{\alpha}{2}} \rho(s) ds , \mbox{\ \  } 0 < r \leq R \medskip\\
     \displaystyle  \frac{2\gamma}{r} , \qquad r > R
\end{array}\right.
\label{eqn:RawGravitySol}
\end{equation}
where
\begin{equation}
\label{eqn:gamma}
\gamma= \frac{G}{(1-\frac{\alpha}{2})} \int_0^R 4 \pi s^2 \rho(s) ds =   \frac{GM}{(1-\frac{\alpha}{2})}
\end{equation}
 Here $M$ is the total matter mass.    As well the middle term in  (\ref{eqn:RawGravitySol})
also has a $1/r^{\alpha/2}$ inflow-singularity, but whose strength is mandated by the matter density, and is absent when $\rho(r)=0$ everywhere. This is a minimal attractor or  ``black hole"\footnote{The term ``black hole" refers to the existence of an event horizon, where the in-flow speed reaches $c$, but otherwise has no connection to the putative ``black holes" of GR.}, and is present in all matter systems.    For the region outside the sun, $r>R$, Keplerian orbits are known to well describe the motion of the planets within the solar system, apart from some small corrections, such as the Precession of the Perihelion of Mercury, which follow from relativistic term in  (\ref{eqn:3spacedynamics}).   The sun, as well as the earth,  has only an  induced  ``minimal attractor'", which affects the interior density, temperature and pressure profiles \cite{Sun}.  These minimal black holes contribute to the external $g=GM^\star/r^2$ gravitational acceleration, through an effective mass $M^\star = M/(1-\alpha/2)$. The 3-space dynamics contributes an effective mass \cite{Book}
\begin{equation}
M_{BH}= \frac{M}{1-\frac{\alpha}{2}}-M=\frac{\alpha}{2} \frac{M}{1-\frac{\alpha}{2}}\approx \frac{\alpha}{2} M
\label{eqn:BHmass}\end{equation}
These induced  black hole ``effective" masses have been detected in numerous  globular clusters and  spherical galaxies and  their predicted effective masses have been confirmed in some 19 such cases, as shown in Fig.\ref{fig:BlackHoles},  \cite{CahillBH2}.   The non-Newtonian effects in (\ref{eqn:RawGravitySol}) are also detectable in bore hole experiments.

\subsection{Earth Bore Holes Determine $\alpha$}
\begin{figure}[t]
\hspace{45mm}\includegraphics[scale=0.18]{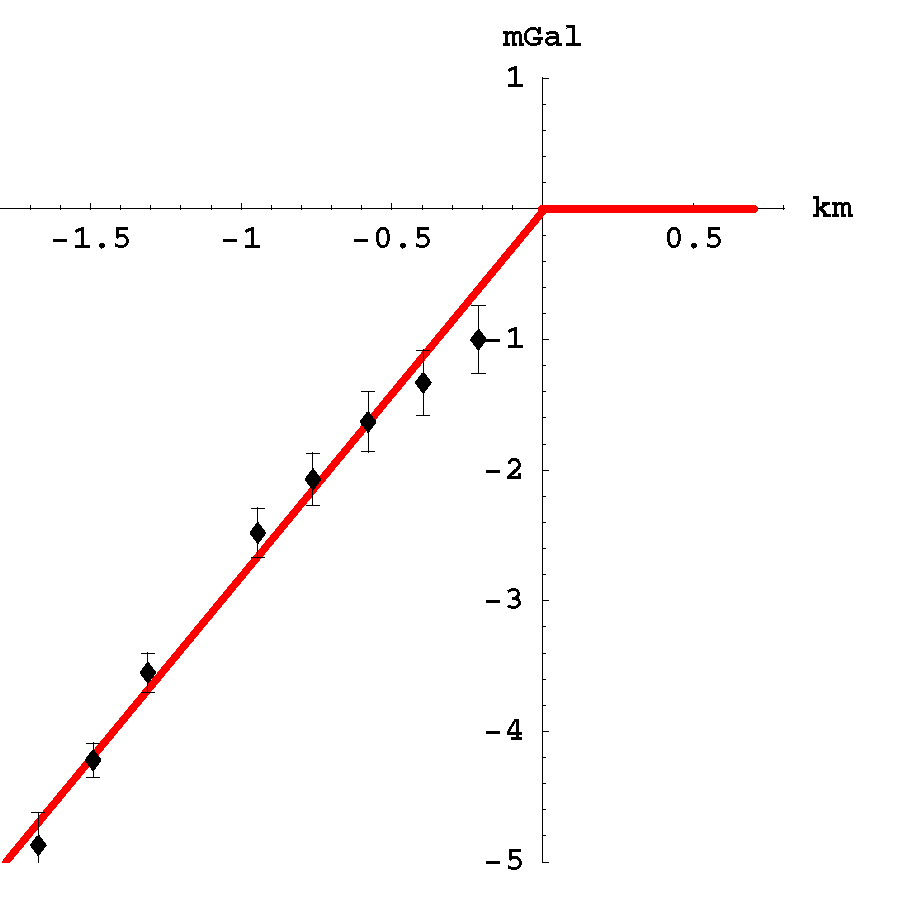}
\caption{\small{The data shows the gravity residuals for the Greenland Ice Shelf \cite{Ander89} Airy measurements of
the $g(r)$  profile,  defined as $\Delta g(r) = g_{Newton}-g_{observed}$, and measured in mGal (1mGal $ =10^{-3}$ cm/s$^2$) and   plotted against depth in km. The borehole effect is that Newtonian
gravity and the new theory differ only beneath the surface, provided that the measured above-surface gravity gradient 
is used in  both theories.  This then gives the horizontal line above the surface. Using (\ref{eqn:ganomaly}) we obtain
$\alpha^{-1}=137.9 \pm  5$ from fitting the slope of the data, as shown. The non-linearity  in the data arises from
modelling corrections for the gravity effects of the   irregular sub ice-shelf rock  topography. The ice density is 920 kg/m$^3$. The near surface data shows that the density of the Greenland ice, compressed snow, does not reach its full density until some 250m beneath the surface - a  known effect.}
\label{fig:Greenland}}\end{figure}

\begin{figure}
\hspace{45mm}\includegraphics[scale=0.18]{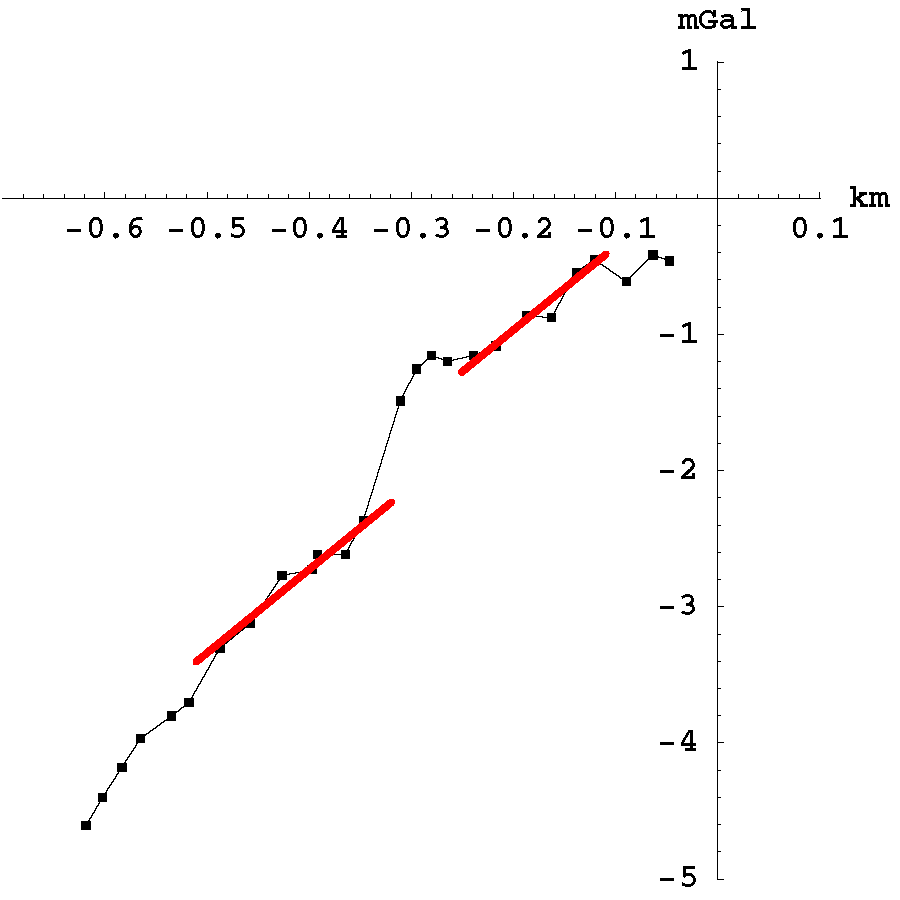}

\hspace{45mm}\includegraphics[scale=0.18]{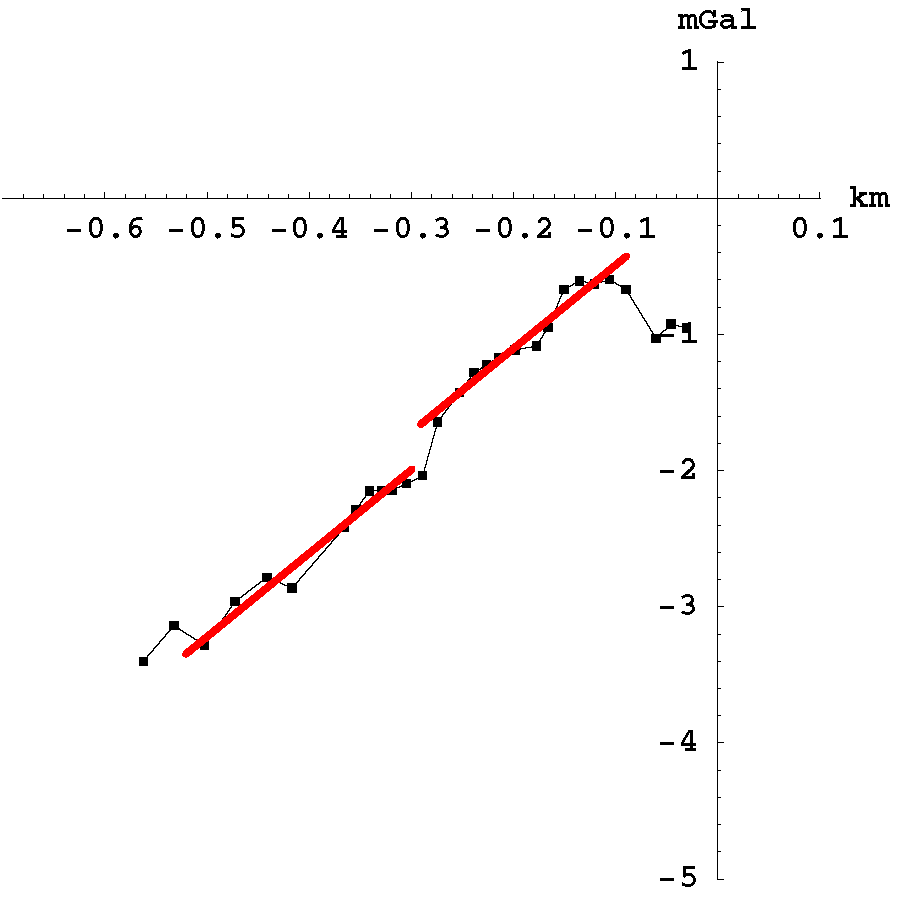}
\caption{\small{  Gravity residuals  $\Delta g(r) $ from two of the Nevada bore hole experiments \cite{Nevada} that give a best fit of $\alpha^{-1}=136.8\pm 3$ on using (\ref{eqn:ganomaly}). Some layering of the rock is evident. The rock density is 2000 kg/m$^3$ in the linear regions.}
\label{fig:Nevada}}\end{figure}

The value of the parameter $\alpha$ in  (\ref{eqn:3spacedynamics}) was first determined from earth bore hole $g$-anomaly data, which shows that gravity decreases more slowly down a bore hole than predicted by Newtonian gravity, see Figs.\ref{fig:Greenland}  and \ref{fig:Nevada}.   From (\ref{eqn:acceleration}) and  (\ref{eqn:RawGravitySol})
we find the gravitational acceleration at radius $r=R+d$ to be
\begin{equation}g(d)=\left\{
\begin{array}{l}
     \displaystyle -\frac{GM}{(1-\alpha/2)(R+d)^2}+\frac{2\pi G\rho(R) d}{(1-\alpha/2)}+... 
     \smallskip  \\    \displaystyle  \mbox{\ \ \ \ \ \  } -\frac{4\pi R^2G \rho(R)G}{(1-\alpha/2)(R+d)^2}     \mbox{\ \ \ \ \  } d<0 
     \medskip \\
     \displaystyle -\frac{GM}{(1-\alpha/2)(R+d)^2}  \mbox{\ \ \  } d>0
\end{array}\right.
\label{eqn:earthg}\end{equation}
In practice the acceleration above the earth's surface must be measured in order to calibrate the anomaly, which defines the coefficient $\overline{GM}=GM/(1-\alpha/2)$ in (\ref{eqn:earthg}). Then the anomaly is
\begin{equation}
\Delta g=g_{NG}(d)\!-\!g(d)=2\pi \alpha G\rho(R) d+O(\alpha^2), \mbox{\ \ \  } d<0
\label{eqn:ganomaly}\end{equation}
to leading order in $\alpha$, and where $g_{NG}(d)$ is the Newtonian gravity acceleration, given the value of $\overline{GM}$ from the above-surface calibration, for a near-surface density  $\rho(R)$. The experimental data then reveals $\alpha$ to be the fine structure constant, to within experimental errors \cite{CahillBH2}. The experiments have densities that differ by more than a factor of  2, so the result is robust.

\begin{figure}
\hspace{40mm}\includegraphics[scale=0.22]{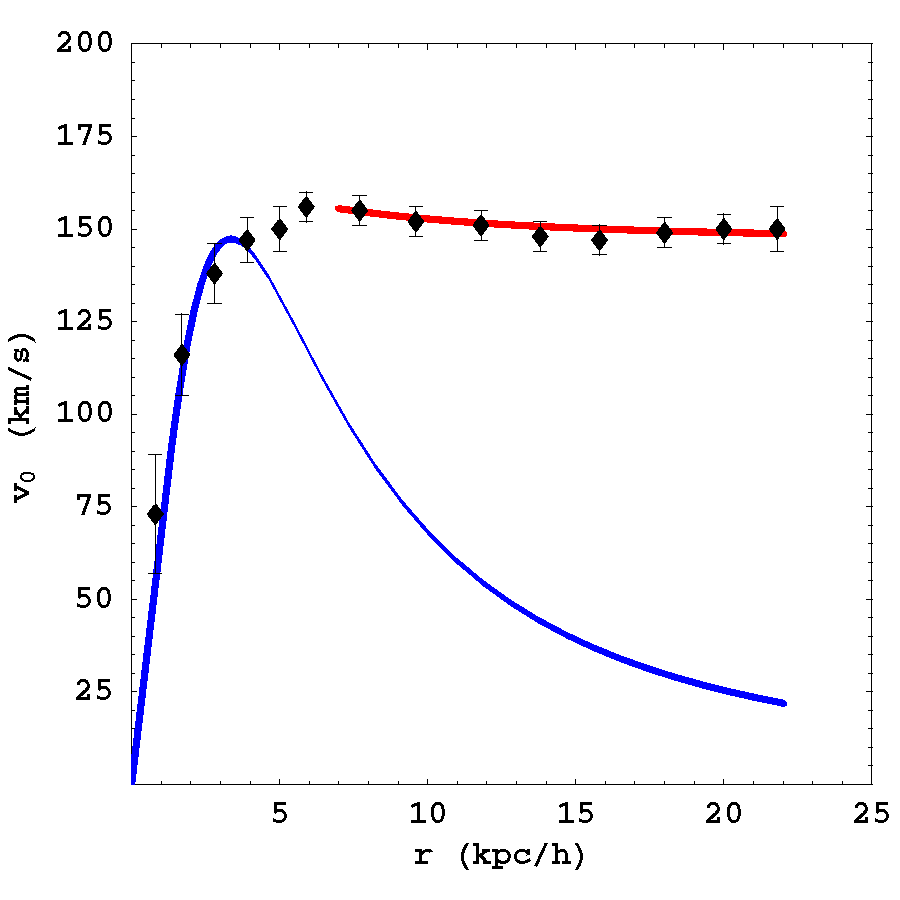}
\caption{\small{ Plots of the rotation speed data for  the  spiral galaxy NGC3198.  Lower curve shows Newtonian gravity prediction, while upper curve shows asymptotic  flat rotation speeds from (\ref{eqn:orbitalspeed}).}}
\label{fig:NGC3198}\end{figure}

\begin{figure}[t]
\hspace{35mm}\includegraphics[scale=0.22]{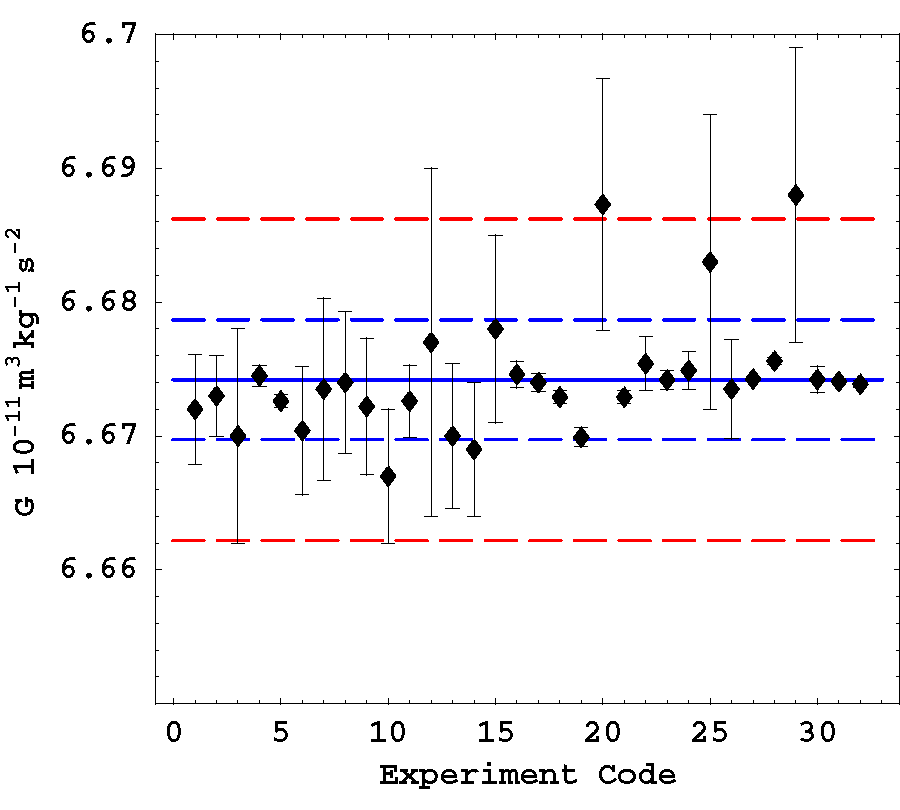}
\caption{\small{Results of precision measurements of $G$ published in the 
last sixty years in which the Newtonian theory was used to analyse the data.  These results 
show  the presence of
a  systematic effect, not in the Newtonian theory, of fractional size $\Delta G/G \approx \alpha/4$.    The
upper horizontal  dashed line shows the value of $G$ from ocean Airy measurements \cite{Ocean}, while the solid   line shows the current CODATA
$G$ value of 6.67428$(\pm0.00067)\times 10^{-11} m^3/kg s^2$, with much lager experimental data range, exceeding  $\pm \alpha G/8$, shown by dashed lines as a guide. The lower horizontal line shows
the actual value of $G$  after removing the space self-interaction effects via $G\rightarrow(1-\alpha/2)G$ from the ocean value of $G$.  The CODATA $G$ value, and its claimed uncertainty, is seen to be spurious.}   
\label{fig:GData}}\end{figure}

\subsection{$G$ Measurement Anomalies}
There has been a long history of anomalies in the measurement of Newton's gravitational constant $G$, see  fig.\ref{fig:GData}.  The explanation is that  the gravitational acceleration external to a piece of matter is only given by application of Newton's inverse square law for the case of a spherically symmetric mass.  For other shapes the $\alpha$-dependent interaction in  (\ref{eqn:3spacedynamics}) results in forces that differ from Newtonian gravity at $O(\alpha)$. The anomalies shown in fig.\ref{fig:GData} result from analysing the one-parameter, $G$,  Newtonian theory, when gravity requires a two parameter, $G$ and $\alpha$, analysis of the data.  The scatter in the measured $G$ values appear to be of $O(\alpha/4)$.  This implies that laboratory measurements to determine $G$ will also measure $\alpha$ \cite{Book}.  

\subsection{Expanding Universe}
\begin{figure}[t]
\hspace{30mm}\includegraphics[scale=0.35]{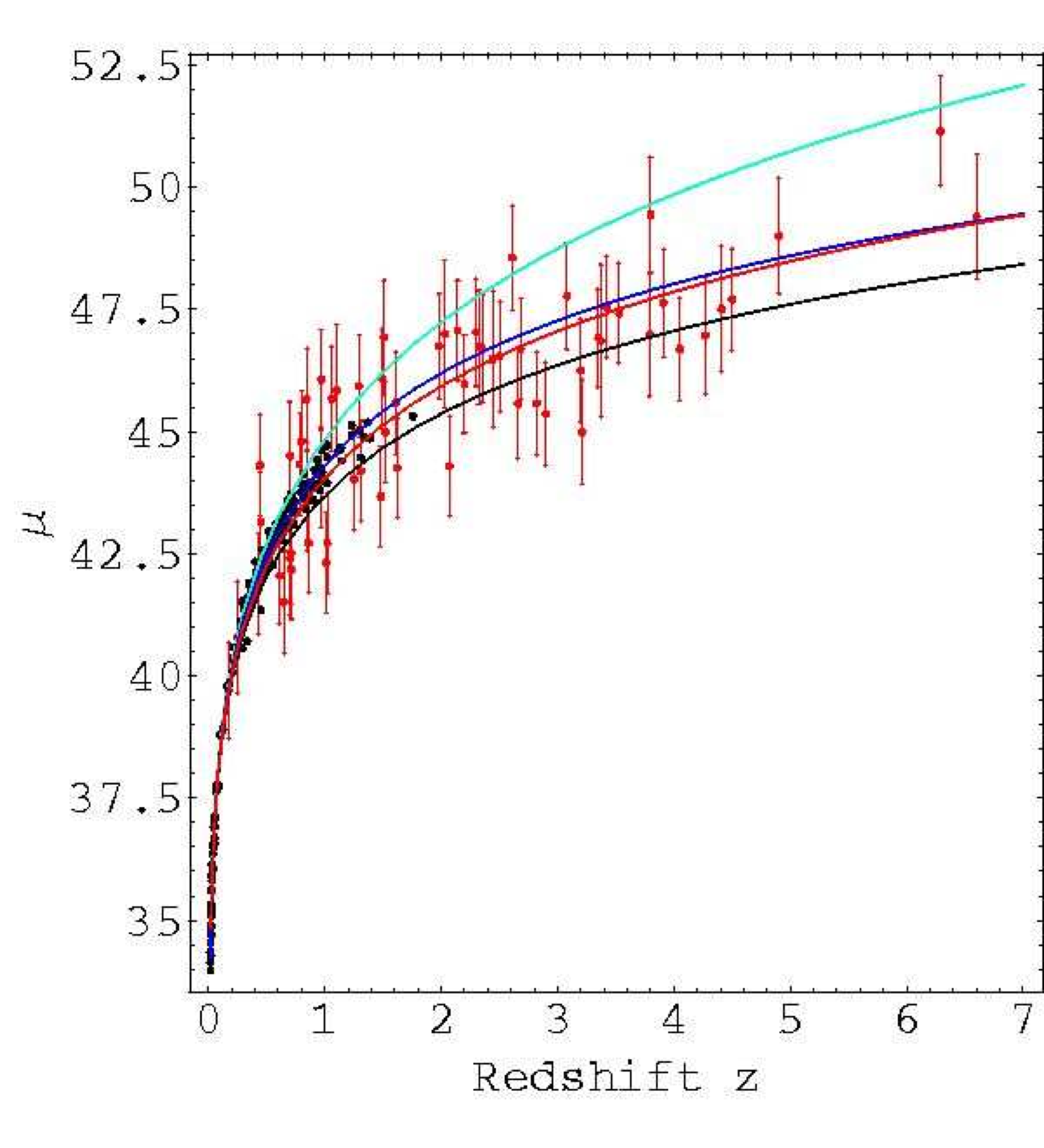}
\vspace{-4mm}\caption{\small{ Hubble diagram showing the supernovae data  using several data sets, and the Gamma-Ray-Bursts data (with error bars).  Upper curve (green)  is $\Lambda$CDM `dark energy' only $\Omega_\Lambda=1$, lower curve  (black) is $\Lambda$CDM matter only $\Omega_M=1$. Two middle curves show best-fit of $\Lambda$CDM `dark energy'-`dark-matter' (blue) and dynamical 3-space prediction (red), and are essentially indistinguishable.   We see that the best-fit  $\Lambda$CDM `dark energy'-`dark-matter' curve essentially converges on the uniformly-expanding parameter-free dynamical 3-space prediction.  The supernova data  shows that the universe is undergoing a uniform expansion, wherein a fit to the FRW-GR  expansion was forced, requiring `dark energy', `dark  matter' and a future `exponentially accelerating expansion'.}
\label{fig:Hubble}}\end{figure}

The dynamical 3-space theory (\ref{eqn:3spacedynamics})  has a time dependent  expanding universe  solution, in the absence of matter,  of the Hubble form
$v(r,t)=H(t)r$ with $H(t)=1/(1+\alpha/2)t$, giving a scale factor   $a(t)=(t/t_0)^{4/(4+\alpha)}$, predicting essentially a uniform expansion rate.  This results in a parameter-free fit to the supernova redshift-magnitude data, as shown in fig.\ref{fig:Hubble},
once the age $t_0=1/H_0$ of the universe at the time of observation is determined  from nearby supernova.  In sharp contrast   the Friedmann model for the universe has a  static solution - no expansion, unless there is matter/energy present.  However to best fit the supernova data  fictitious ``dark matter" and ``dark energy" must be introduced, resulting in the $\Lambda$CDM model.  The amounts $\Omega_{\Lambda}=0.73$ and $\Omega_{DM}=0.23$ are easily determined by best fitting the 
$\Lambda$CDM model to the above uniformly expanding result, without reference to the observational supernova data.  But then the $\Lambda$CDM has a spurious exponential expansion which becomes more pronounced in the future. This is merely a consequence of extending a poor curve fitting procedure beyond the data.  The 3-space dynamics (\ref{eqn:3spacedynamics}) results in a hotter universe in the radiation dominated epoch, with effects on Big Bang Nucleosynthesis \cite{CahillThermal}, and also a later decoupling time of some $1.4\times10^6$ years.

\section{Primordial Black Holes}
In the absence of matter the dynamical 3-space equation  (\ref{eqn:3spacedynamics}) has black hole solutions of the form 
\begin{equation}
v(r) =- \frac{\beta}{r^{\alpha/4}}
\label{eqn:blackhole}\end{equation}
 for arbitrary $\beta$, but only when $\alpha \neq 0$.  This will produce a long range gravitational acceleration, essentially decreasing like $1/r$,
  \begin{equation} g(r)=-\frac{\alpha\beta}{4r^{1+\alpha/2}}\label{eqn:bhg}\end{equation}
  as observed in spiral galaxies.
The inflow in  (\ref{eqn:blackhole}) describes an inflow singularity or ``black hole" with arbitrary strength.
This is unrelated to the putative black holes of General Relativity. This corresponds to a primordial black hole. The dark matter density for these black holes is
\begin{equation}
\rho_{DM}(r)=\frac{\alpha\beta^2(2-\alpha)}{256 \pi G r^{2+\alpha/2}}
\label{eqn:bhdm}\end{equation}
This decreases like $1/r^2$ as indeed determined by the ``dark matter"  interpretation of the flat rotation curves of spiral galaxies. Here, however, it is a purely 3-space self-interaction effect.

In general a spherically symmetric matter distribution may have a static solution which is a linear combination of  the inhomogeneous matter induced solution in (\ref{eqn:RawGravitySol}) and the square of the homogeneous primordial black hole solution in (\ref{eqn:blackhole}), as 
(\ref{eqn:sphericalsym}) is linear in $v(r)^2$ and its spatial derivatives.. However this is unlikely to be realised, as a primordial black hole would cause a precocious in-fall of matter, which is unlikely to remain spherically symmetric, forming instead spiral galaxies.  

\subsection{Spiral Galaxy Rotation Curves}
Spiral galaxies  are formed by matter in-falling on  primordial black hole, leading to rotation of that matter, as the in-fall will never be perfectly symmetric. 
The black hole acceleration in (\ref{eqn:bhg}) would support a circular matter orbit with orbital speed
\begin{equation}
v_o(r)=\frac{(\alpha\beta)^{1/2}}{2r^{\alpha/4}}
\label{eqn:orbitalspeed}\end{equation}
which is the observed asymptotic ``flat" orbital speed in spiral galaxies, as illustrated in Fig.\ref{fig:NGC3198} for the spiral galaxy NGC3198.  So the flat rotation curves are simply explained by (\ref{eqn:3spacedynamics}).

\section{Primordial Filaments}
\begin{figure}
\vspace{40mm}
\hspace{30mm}
\setlength{\unitlength}{0.9mm}
\hspace{0mm}\begin{picture}(0,0)
\thicklines

\definecolor{hellgrau}{gray}{.8}
\definecolor{dunkelblau}{rgb}{0, 0, .9}
\definecolor{roetlich}{rgb}{1, .7, .7}
\definecolor{dunkelmagenta}{rgb}{.9, 0, .0}
\definecolor{green}{rgb}{0, 1,0.4}
\definecolor{black}{rgb}{0, 0, 0}

\color{dunkelmagenta}
\put(20,0){\line(0,1){40}}
\put(24,0){\line(0,1){40}}
  \qbezier(20,40)(22,43)(24,40)
    \qbezier(20,40)(22,37)(24,40)
        \qbezier(20,00)(22,-3)(24,0)
        \put(21,0){\circle{2}}  \put(22,1){\circle{2}}  \put(23,0){\circle{2}}
          \put(21,3){\circle{2}}  \put(22,2){\circle{2}}  \put(23,3){\circle{2}}
                \put(21,4){\circle{2}}  \put(22,6){\circle{2}}  \put(23,5){\circle{2}}
                  \put(21,7){\circle{2}}  \put(22,7){\circle{2}}  \put(23,8){\circle{2}}
                      \put(21,10){\circle{2}}  \put(22,9){\circle{2}}  \put(23,9){\circle{2}}
                         \put(21,13){\circle{2}}  \put(22,12){\circle{2}}  \put(23,14){\circle{2}}
                            \put(21,15){\circle{2}}  \put(22,17){\circle{2}}  \put(23,16){\circle{2}}
                                 \put(21,20){\circle{2}}  \put(22,19){\circle{2}}  \put(23,19){\circle{2}}
                                     \put(21,24){\circle{2}}  \put(22,22){\circle{2}}  \put(23,23){\circle{2}}
                                         \put(21,26){\circle{2}}  \put(22,27){\circle{2}}  \put(23,25){\circle{2}}
                                                \put(21,28){\circle{2}}  \put(22,30){\circle{2}}  \put(23,29){\circle{2}}
                                                      \put(21,30){\circle{2}}  \put(22,33){\circle{2}}  \put(23,31){\circle{2}}
                                                        \put(21,33){\circle{2}}  \put(22,35){\circle{2}}  \put(23,33){\circle{2}}
                                                          \put(21,35){\circle{2}}  \put(22,37){\circle{2}}  \put(23,36){\circle{2}}
                                                               \put(21,40){\circle{2}}  \put(22,39){\circle{2}}  \put(23,39){\circle{2}}
          
        \color{dunkelblau}
        \put(22,40){\vector(0,1){7}}
            \put(24,44){\large z}
        \put(22,20){\line(1,0){36}}
              \put(60,19){\large r}
         \put(22,20){\line(2,-1){20}}
          \qbezier(42,10)(52,18)(52.5,20)
          
           \put(22,10){\line(1,0){10}}
         \put(22,10){\line(2,-1){20}}
          \qbezier(42,0)(54,8)(52.5,10)
          
                 \put(38,4){\Large $d$}
                             \put(30,16.26){\large $\theta$}
          
          \put(42,10){\line(0,-1){10}}
                    \put(52.5,10){\line(0,1){10}}
                            \put(32,15){\line(0,-1){10}}
                                    \qbezier(32,15)(40,18)(40,20)
                                    \put(57,6){\vector(-3,1){10}}
               \put(57,7){\Large$v, {\bf g}$}   
               
                 \put(30,-1){\Large$R_1$}  \put(40,-5){\Large$R_2$}            

              \end{picture}
\vspace{4mm} \caption{\small{ Sector integration volume, with radii $R_1$ and $R_2$,  about a filament. For the filament to exist the quantum foam substructure to 3-space must be invoked at short distances.}}
 \label{fig:Filament}
\end{figure}
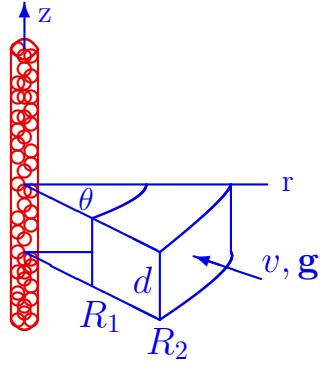

Eqn.(\ref{eqn:3spacedynamics}) also has cosmic filament solutions.  Writing  (\ref{eqn:3spacedynamics}) in cylindrical coordinates $(r,z,\phi)$, and assuming cylindrical symmetry  with translation invariance along the $z$ axis, we have for a radial flow $v(r,t)$
\begin{equation}
\frac{1}{r}\frac{\partial v}{\partial t}+\frac{\partial v^\prime}{ \partial t}+\frac{vv^\prime}{r}+v^{\prime 2}+vv^{\prime\prime}+
\alpha\frac{vv^\prime}{4r}=0
\label{eqn:filaments}\end{equation}
where here the radial distance $r$ is the distance perpendicular to the $z$ axis. This has static solutions
with the form
\begin{equation}
v(r) =- \frac{\mu}{r^{\alpha/8}}
\label{eqn:filamentsol}\end{equation}
for arbitrary $\mu$. The gravitational acceleration is   long-range and attractive to matter, i.e. ${\bf g}$ is directed inwards towards the filament,
\begin{equation}
g(r) = -\frac{\alpha\mu^2}{8r^{1+\alpha/4}}
\label{eqn:filamentg}\end{equation}
This is for a single infinite-length filament.
The dark matter density (\ref{eqn:darkmatter})  is
\begin{equation}
\rho_{DM}(r)=-\frac{\alpha \mu^2 }{1024  \pi G r^{2+\alpha/4}}
\label{eqn:filamentdm}\end{equation}
and negative. But then (\ref{eqn:NGplus}), with $\rho=0$, would imply a repulsive matter acceleration by the filament, and  not attractive as in 
(\ref{eqn:filamentg}).  To resolve this we consider the sector integration volume in Fig.\ref{fig:Filament}.  
We obtain from (\ref {eqn:filamentg})  and using the divergence theorem (in which ${\bf dA}$ is directed outwards from the integration volume)
\begin{equation}
\int_{\cal V}\! \nabla\!\cdot\!{\bf g}dv\!= \!\int _{\cal A}\!{\bf g}\!\cdot\! {\bf dA}=\frac{\alpha \mu^2 \theta d }{8}\left( \frac{1}{R_1^{\alpha/4}}-\frac{1}{R_2^{\alpha/4}}\right) 
\label{eqn:gdiv}\end{equation}
which is positive because $R_1 < R_2$. This is consistent with    (\ref{eqn:NGplus}) for the negative $\rho_{DM}$, but only if $R_1$ is finite.  However if $R_1=0$, as for the case of the integration sector including the filament axis,  there is no $R_1$ term in (\ref{eqn:gdiv}), and the integral is now negative.  This implies that  (\ref{eqn:filamentsol}) cannot be the solution for some small $r$. The filament solution is then only possible if  the dynamical 3-space equation (\ref{eqn:3space}) is applicable only to macroscopic distances, and at short distances higher order derivative terms become relevant, such as $\nabla^2(\nabla \cdot{\bf v})$.  Such terms indicate the dynamics of the underlying quantum foam, with (\ref{eqn:3space})  being  a derivative expansion, with higher order dervatives becoming more significant at shorter distances.

\section{Filament Gravitational Lensing}
We must   generalise the Maxwell equations so that the electric and magnetic  fields are excitations within the dynamical 3-space, and not of the embedding space.  The minimal form in the absence of charges and currents is 
 \begin{eqnarray}
\displaystyle{ \nabla \times {\bf E}}&=&\displaystyle{-\mu_0\left(\frac{\partial {\bf H}}{\partial t}+{\bf v.\nabla H}\right)},
 \mbox{\ \ \ }\displaystyle{\nabla.{\bf E}={\bf 0}},  \nonumber \\
 \displaystyle{ \nabla \times {\bf H}}&=&\displaystyle{\epsilon_0\left(\frac{\partial {\bf E}}{\partial t}+{\bf v.\nabla E}\right)},
\mbox{\ \ \  }\displaystyle{\nabla.{\bf H}={\bf 0}}\label{eqn:E18}\end{eqnarray}
which was first suggested by Hertz in 1890  \cite{Hertz}, but with ${\bf v}$ then being only a constant vector field. As easily determined  the speed of EM radiation is now $c=1/\sqrt{\mu_0\epsilon_0}$ with respect to the 3-space.
The time-dependent and inhomogeneous  velocity field causes the refraction of EM radiation. This can be computed by using the Fermat least-time approximation. This ensures that EM waves along neighbouring paths are in phase. Then the EM ray paths  ${\bf r}(t)$ are determined by minimising  the elapsed travel time: 
\begin{equation}
T=\int_{s_i}^{s_f}\frac{ds\displaystyle{{ |} \frac{d{\bf r}}{ds}|}}{|c\hat{{\bf v}}_R(s)+{\bf v}(\bf{r}(s),t(s)|},
\label{eqn:EMtime}\end{equation}
\begin{equation}{\bf v}_R=  \frac{d{\bf r}}{dt}-{\bf v}(\bf{r}(t),t)
\label{eqn:lighttime}\end{equation}
by varying both ${\bf r}(s)$ and $t(s)$, finally giving ${\bf r}(t)$. Here $s$ is a path parameter, and $c\hat{\bf v}_R$ is the velocity of the EM radiation wrt the local 3-space, namely $c$. The denominator in (\ref{eqn:EMtime}) is the speed of the EM radiation wrt the observer's Euclidean spatial coordinates.
Eqn.(\ref{eqn:EMtime}) may be used to calculate the gravitational lensing by black holes, filaments and by ordinary matter, using the appropriate  3-space velocity field. Because of the  long-range nature of the inflow for black holes and filaments, as in (\ref{eqn:blackhole})  and (\ref{eqn:filamentsol}), they produce strong lensing, compared to that for ordinary matter\footnote{Eqn:(\ref{eqn:EMtime}) produces the known sun light bending \cite{Review}.}, and also compared with the putative black holes of GR, for which the in-flow speed decreases like $1/\sqrt{r}$, corresponding to the acceleration field decreasing like $1/r^2$.  The EM lensing caused by filaments and black holes is the basis of the stochastic tomographic technique for detecting these primordial 3-space structures.

\section{Filament and Black Hole Networks}
The dynamical 3-space equation produces analytic solutions for the cases of a single primordial black hole, and  a single, infinite length, primordial filament. This is because of the high symmetry of theses cases.  However analytic solutions corresponding to a network of  finite length filaments joining at black holes, as shown in Fig.\ref{fig:Filaments}, are not known.  For this case numerical solutions will be needed.   It is conjectured that the network is a signature of primordial imperfections or defects from the epoch when the 3-space was forming, in the earliest moments of the big bang. It is conjectured that the network of filaments and black holes form a cosmic network of  sheets and voids.  This would amount to a dynamical breakdown of the translation invariance of space. Other topological defects are what we know as quantum matter \cite{Book}.

\section{Conclusions} 
The recent discovery that a dynamical 3-space exists has resulted in  a comprehensive  investigation of the new physics, and which has been checked against numerous experimental and observational data. This data ranges from laboratory Cavendish-type $G$ experiments to the expansion of the universe which, the data clearly shows, is occurring at a uniform rate, except for the earliest epochs.  Most significantly the dynamics of space involves two parameters: $G$, Newton's gravitational constant, which determines the rate of dissipative flow of space  into matter, and $\alpha$, which determines the space self-interaction dynamics.  That this is the same constant that determines the strength of electromagnetic interactions shows that a deep unification of physics is emerging.  It is the $\alpha$ term in the  space dynamics that determines almost all of the new phenomena. Most importantly the epicycles of spacetime physics, {\it viz} dark matter and dark energy, are dispensed with.

\end{document}